\documentclass{optica-article}

\journal{opticajournal} 
\usepackage{amsmath}
\usepackage{xcolor}
\usepackage{empheq}
\articletype{Research Article}

\begin{document}

\title{Single mode lasing and spectral narrowing in photonic crystal line-defect cavities via spatially selected Bloch modes}

\author{SHU-NING DING,\authormark{1,2,*}  LING-FANG WANG,\authormark{1,2}, XIAO-TIAN CHENG,\authormark{1,2}, JIA-WANG YU,\authormark{1,2} DAI-BAO HOU,\authormark{1,2} YI-FENG LIU,\authormark{1,2} ZHE FENG,\authormark{1,2} YI ZHAO,\authormark{1,2} YANG-CHEN ZHENG,\authormark{1,2} XING LIN,\authormark{1,2} FENG LIU,\authormark{1} CHEN-HUI LI,\authormark{1,2,3} and CHAO-YUAN JIN\authormark{1,2,4,\dag}}

\address{\authormark{1} College of Information Science and Electronic Engineering, Zhejiang University, Hangzhou 310027, China\\
\authormark{2} State Key Laboratory of Silicon and Advanced Semiconductor Materials, ZJU-Hangzhou Global Scientific and Technological Innovation Center, Zhejiang University, Hangzhou 311200, China\\
\authormark{3} Research Institute of Intelligent Computing,
Zhejiang Lab,
Hangzhou 311121, China\\
\authormark{4} College of Integrated Circuits, Zhejiang University, Hangzhou, Zhejiang 311200, China}

\email{\authormark{*}dingshn@zju.edu.cn}
\email{\authormark{\dag}jincy@zju.edu.cn} 


\begin{abstract*} 
The demand for high-efficiency and miniaturized on-chip light sources drives continuous innovation in photonic crystal (PhC) microcavity lasers. The presence of slow-light effects in PhC microcavities leads to the mode competition between Bloch modes resulting in multi-mode lasing, which obstructs the dense integration of PhC lasers. Here, we theoretically verify a technical scheme for the single-mode lasing of PhC line-defect-cavity lasers by spatially pumping a certain Bloch mode via optical interference.We demonstrate the capability to select a specific longitudinal mode to lase with a side mode suppression ratio (SMSR) exceeding 30 dB. The interaction between optical interference fringes and the vacuum electromagnetic field inside the PhC cavity improves the linewidth and noise characteristics of lasers. This scheme of Bloch mode selection provides a novel and viable tool for the manipulation of PhC microcavity lasers.
\end{abstract*}

\section{Introduction}
As the bandwidth demand for interchip communication is continuously rising, monochromatic light sources with high modulation speed and small device footprint become increasingly important for the miniaturization of photonic integration circuits \cite{Shi:22,Wang2020b}. Photonic crystal (PhC) microcavity lasers, which feature both high quality factor and small mode volume, are promising candidates for dense integration with high performance, positioning themselves as ideal light sources for on-chip photonic integration \cite{Takeda:21,10.1117/1.AP.1.1.014002}. However, PhC cavities which possess micrometer-size carrier reservoirs generally come with severe gain saturation and mode competition in lasing \cite{https://doi.org/10.1002/lpor.202400218}. A great deal of nanophotonic approaches have been proposed to achieve effective mode manipulation in PhC lasers by introducing distributed feedback gratings \cite{10.1063/1.2716972}, coupled cavities \cite{10.1063/1.1427158,Wang:22,Burgwal2023,PhysRevA.91.063807}, broken PT symmetry \cite{Wang:21}, or spatial injection \cite{Kim2020}. 

Besides, a quantum optical approach based on the manipulation of vacuum electromagnetic field has been recently proposed for the mode selection in Fabry-Pérot-like microcavities which does not require special design of the cavity structure \cite{Wang:20}. A SMSR exceeding 40 dB has been predicted in theory while the spatial profile of optical interference perfectly matches the field distribution of a selected lasing mode. However, this approach requires a spatial resolution at or beyond the wavelength scale to fit the mode pattern, making such precise wavelength modulation experimentally difficult to implement. It has been only observed that single mode lasing occurs in microbottle lasers when one of the stripes in the interference pattern precisely overlaps with the equator of the microbottle cavity\cite{Gu2017}. 

Alternatively, the constraint on the spatial resolution might be relaxed in PhC line-defect microcavities, because the presence of periodicity in PhCs modifies the electromagnetic field to form Bloch waves, which have large nodes compared with the lasing wavelength. A set of low-order FP modes can be constructed simultaneously, which satisfy the round-trip phase relation at FP resonances and share the small group velocity of Bloch waves \cite{10.1063/1.5012112}. To achieve mode selection in PhC lasers, one needs to consider not only the interference patterning for lasing but also the spatial nature of Bloch modes in the slow-light region, which hence provides new insights on the mode selection in nanophotonic lasers. 

In this work, the mode selection in PhC lasers is tested on a Bloch-mode scenario. Spatial injection based on optical interference with a slow-varying envelope is installed on line-defect PhC cavities. The mode profiles of Bloch modes are taken into account in a traveling wave rate equation model, whereas lasing modes are closely packed in the spectrum due to the periodic lattice structure and slow-light effect in PhC cavities. Although a high overlap between the optical interference pattern and the field distribution of a lasing mode indeed improves mode selection \cite{Cheng2024QuantumVacuum}, we neglect the wavelength-scale change of the vacuum field by exploring the slow-varying envelope of Bloch modes. By optimizing the geometric configuration of the optical interference, any one of the three longitudinal modes can be chosen to lase with a SMSR higher than 30 dB with a reduced noise level and narrowed lasing linewidth.

\section{Theoretical analysis and modeling}

\subsection{Distribution of Bloch modes in photonic crystal microcavities}
\begin{figure}[h]
\centering\includegraphics[width=7cm]{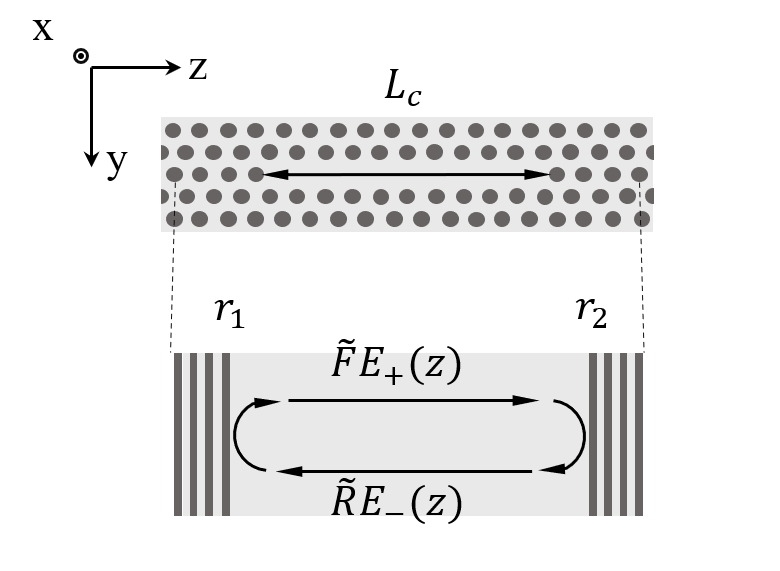}
\caption{The propagation of forward and backward Bloch waves in a PhC microcavity and the reflection at both ends of the periodic structure.}
\label{fig:1}
\end{figure}

By eliminating $N$ air holes from a designated row in a PhC structure, a line-defect cavity or a so-called $LN$ cavity is created. A plane wave expansion method is utilized to describe Bloch modes \cite{Okano_2010}, 
\begin{equation}
\mathbf{E_{-}}(\mathbf{r})= e^{ik_z \cdot \mathbf{r} } \mathbf{e_+} (\mathbf{r}),
\label{eq:1} 
\end{equation}
\begin{equation}
\mathbf{E_{+}}(\mathbf{r})= e^{-ik_z \cdot \mathbf{r}+i\theta } \mathbf{e_-} (\mathbf{r}),
\label{eq:2} 
\end{equation}
where $\mathbf{E_{+}}(\mathbf{r})$ and $\mathbf{E_{-}}(\mathbf{r})$ indicate the forward- and reverse-propagating electric field, respectively, $\theta$ represents the phase difference between the two sets of plain waves with the opposite direction of propagation ($0 \leq \theta \leq 2\pi$). The wave vector $k_z(\omega)$ is defined by the dispersion relation of the PhC microcavity along the $z$ axis, and $\mathbf{e_{\pm}}(\mathbf{r}) $ is a periodic function decided by the lattice constant, ${a}$ \cite{Vasco2018}. Considering the reflection at the end of a one-dimensional PhC cavity, the field distribution in an $LN$ cavity can be written as the superposition of two opposite Bloch modes \cite{saleh2019fundamentals},
\begin{equation}
E(z,t) = \left[ \tilde{F} E_+ (z) + \tilde{R} E_- (z) \right] e^{i\omega t},
\label{eq:3} 
\end{equation}
where $\tilde{F}/\tilde{R}$ is the amplitude of the forward / backward Bloch mode. Only the lateral field along the cavity centerline is analyzed to illustrate resonant mode characteristics as shown in Fig.\ref{fig:1}. For a one-dimensional cavity of length $L_c$, we set the point $z=0$ at the left end of the cavity. The boundary conditions read,  
\begin{equation}
F(0,t) E_+ (0) r_1 + R(0,t) E_- (0) = 0,
\label{eq:4} 
\end{equation}
\begin{equation}
F(L_c,t) E_+ (L_c) + R(L_c,t) E_- (L_c) r_2 = 0,
\label{eq:5} 
\end{equation}
where $r_{1}$ and $r_{2}$ are the reflectivities close to 1 \cite{lalanne2008photon} of the PhC cavity’s left and right facets.

After substituting equations \eqref{eq:1} and \eqref{eq:2}, we can obtain,
\begin{equation}
1+e^{i\theta} = 0,
\label{eq:6} 
\end{equation}
\begin{equation}
e^{ik_zL_c}+e^{-i k_z L_c+i\theta} = 0,
\label{eq:7} 
\end{equation}
where $\theta$ and $k_{z}$ can be solved for $\theta=\pi,k_{z}=p\pi/L_c$, where $p$ is a positive integer. The wave vector of Bloch mode at the order $m$ in the $LN$ cavity is a real number $k_m  =\pi/a+m\pi/L_c$, where $\pi/a$ is the magnitude of the wave vector at the boundary of Brillouin zone \cite{nano11113030}. Because of the dispersive nature of the line-defect photonic crystal waveguide, the group velocity near the edge of Brillouin zone is significantly reduced. As a result, Bloch modes with wave vectors close to $\pi/a$ interact with the gain medium for a longer duration, making lower-order Bloch modes more likely to reach the lasing threshold. The fundamental mode (m = 1) exhibits minimal detuning from the band edge.
The field of the resonant mode in real space can be represented as the product of a slow-varying envelope and a fast-varying periodic function,
\begin{equation}
E_m(z) \propto \underbrace{\sin\left(\frac{\pi}{a} z\right) }_{\text{Fast-varying wave}} \cdot\underbrace{\sin\left(\frac{m\pi }{L_c} z\right)}_{\text{ Slow-varying envelope}}, \quad 0\leq z \leq L_c,
\label{eq:8} 
\end{equation}
where $a\ll L_c$. This is equivalent to loading a low-frequency envelope with a period of $2L_c/m$ onto a standing wave distribution with a period of $2a$, which is different from the mode field inside a classical FP cavity where the field amplitude is almost independent of the position within the cavity. For the fundamental Bloch mode ($m=1$), the envelope of the electric field distribution has only one maximum, for the second Bloch mode ($m=2$), it has two maxima, and so forth. In experiments, Bloch modes can be distinguished on the basis of the spatial distribution of the slow-varying envelope with a node size much larger than the lasing wavelength.

\subsection{Interference pumping based on the slow-varying envelope of Bloch modes}
\begin{figure}[h]
\centering\includegraphics[width=10cm]{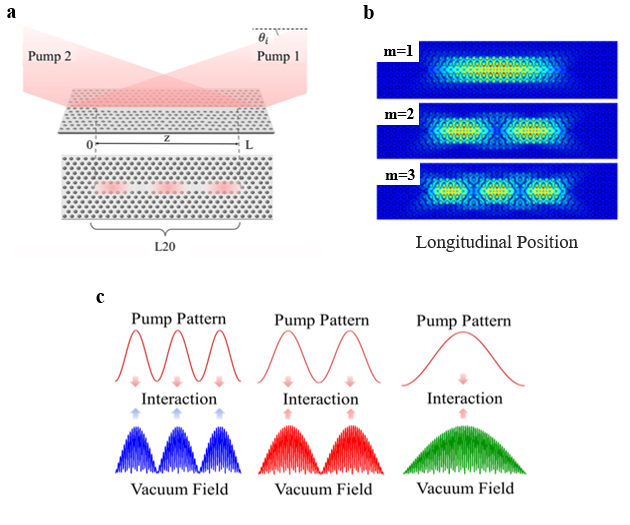}
\caption{ Schematic for optical interference pumping and the mode selection in a PhC cavity. a. Interference pumping on an $L20$ PhC cavity. b. The field energy distributions of the first three Bloch modes in the $L20$ cavity obtained via finite element method.} c. The interaction between the vacuum electromagnetic field of three Bloch modes in the $L20$ cavity and the external injection which sketches the slow-varying profile of Bloch modes.
\label{fig:2}
\end{figure}

A theoretical model based on one-dimensional wave propagation is developed to describe an optical interference pumped $LN$ PhC microcavity laser. Figure \ref{fig:2}(a) shows a schematic of two coherent light beams incident at the same angle from both sides of a PhC microcavity. 
The power distribution of optical interference, $S_{in}(z)$, projected onto the PhC line-defect cavity expresses as,
\begin{equation}
S_{in}(z) = 2A^2 - 2A^2 \cos\left(2k_i z \sin\theta_i\right),
\label{eq:9} 
\end{equation}
where $A$ is the amplitude of electric field, $k_{i}$ represents the wave number of interference light, $z$ represents the longitudinal position, and $\theta_{i}$ stands for the incident angle.

Using the expression for the mode field in PhC line-defect cavities, we distinguish different Bloch modes by extracting the number of nodes in the slow-varying envelope. Assuming that the intensity of the mode field at antinode positions is $A_{0}$, the intensity distribution of the Bloch modes expresses as,
\begin{equation}
I =\ A_0^2 | \underbrace{\sin\left(\frac{\pi}{a} z\right) }_{\text{Fast-varying wave}} \cdot\underbrace{\sin\left(\frac{m\pi}{L_c} z\right)}_{\text{ Slow-varying envelope}}|^2.
\label{eq:10} 
\end{equation}

The eigenfrequencies of the LN cavity were simulated using the finite element method to extract the Bloch mode field distributions, which were then compared with the analytically derived results. Taking the first three Bloch modes as examples, Figure \ref{fig:2}(b) shows the normalized electric field energy distribution $|E_m|^2$. The field energy distribution along the central axis of the cavity in Fig.\ref{fig:2}(b) closely matches the vacuum field shown in Fig.\ref{fig:2}(c).

According to Equations \eqref{eq:9} and \eqref{eq:10}, the period of the interference pattern can be controlled by varying the angle of incidence and the wavelength of interfering beams, to precisely overlap with the slow-varying envelope of Bloch modes inside the cavity. Interference fringes with a period comparable to the cavity length and at the order of micrometers are employed, which provides a feasible method for the practical realization of single-mode PhC lasers via spatial pumping.

\subsection{ Traveling wave rate equation}
Traveling-wave rate equations are employed to analyze lasing properties of the spatially pumped PhC cavity. Using Equation \eqref{eq:3}, we represent the electric field distribution of the lasing light as a superposition of two oppositely propagating waves. The mode field amplitude $F$ and $R$ satisfy time-dependent traveling wave equations by neglecting the coupling between forward and backward optical fields \cite{1291710,Hantschmann:18,1291704},
\begin{equation}
\frac{dF(t,z)}{dt} \cdot \frac{1}{v_g} + \frac{dF(t,z)}{dz} = \left( \Gamma g(t,z,\lambda) P(z) F_p - \alpha - \frac{\alpha_{MIR}}{S} \right) F(t,z) + S_F (t,z)
\label{eq:11}
\end{equation}

\begin{equation}
\frac{dR(t,z)}{dt} \cdot \frac{1}{v_g} - \frac{dR(t,z)}{dz} = \left( \Gamma g(t,z,\lambda) P(z) F_p - \alpha - \frac{\alpha_{MIR}}{S} \right) R(t,z) + S_R (t,z)
\label{eq:12} 
\end{equation}
where $v_{g}$ is the group velocity, $g$ is the material gain, $\Gamma$ is the confinement coefficient factor, $P$ is the effective factor governed by the spatial distribution of vacuum field, $F_{P}$ is Purcell factor and $\alpha$ is the internal loss. Slow-down factor, $S = n_{g}^{phc}/n_{g}^{b}$, is introduced by the slow light effect which averages the mirror loss $\alpha_{MIR}$ \cite{PhysRevLett.116.063901}. Here, $ n_{g}^{phc}$ is the group index of line-defect waveguide without active material and $n_{g}^{b}$ is the refractive index of the cladding slab. $S_{F}$ and $S_{R}$ represent the forward and reverse field of the spontaneous emission coupled into the lasing mode. The spatial factor of vacuum field is simplified to the following expression by using the rotation wave approximation,
\begin{equation}
P(z) = \left| \sin\left(\frac{\pi }{a} z\right)\cdot \sin\left(\frac{m\pi }{L_c} z\right)   \right|.
\label{eq:13} 
\end{equation}

According to Fermi's golden rule \cite{10.1143/ptp/5.4.570}, the change in the local density of the photon states of the vacuum field changes the radiation rate of light sources, modulating both spontaneous and stimulated emission \cite{Jin2014,10.1063/1.3697702}. The degree of enhancement can be expressed by the Purcell factor $F_{p}$ \cite{Zhou2020},
\begin{equation}
F_P(\lambda) = \frac{3Q} {4\pi V} (\frac{\lambda} n)^3,
\label{eq:14} 
\end{equation}
where $Q$ is the quality factor, $V$ is the volume of the cavity mode and $\lambda$ is the mode wavelength. Purcell effect indicates that the spontaneous emission characteristics of a radiative source are determined not only by intrinsic properties, but also by the electromagnetic environment \cite{Purcell1995}. As shown in Fig.\ref{fig:2}(c), the proposed single-mode operation depends on the spatial overlap between the vacuum field distribution of Bloch modes and the optical interference pumping.

In the simulation, a parabolic approximation is used to describe the material gain \cite{1071656},
\begin{equation}
g(t,z,\lambda) = \frac{g_{N_c} \left( {N_c}(t,z) - N_0 \right) - \left( \frac{\lambda_0 - \lambda}{G_0} \right)^2}{2 \left( 1 + \varepsilon N_p (t,z) \right)}
\label{eq:15} 
\end{equation}
where $g_{N}$ is the differential gain, $N_{0}$ represents the transparent carrier density, $N_c$ is the carrier density, $G_{0}$ stands for the parabolic gain fitting factor, and $\varepsilon$ denotes the gain compression factor when gain saturation occurs. Photon density $N_{p}$  is calculated by, 
\begin{equation}
N_{p} (t,z) = \frac{\varepsilon_0 |E(t,z)|^2 }{h \nu}
\label{eq:16} 
\end{equation}
where $\varepsilon_{0}$ is the vacuum dielectric constant, and $h$ is the Planck constant. The total energy of the optical field at a given frequency $\nu$ is denoted $\varepsilon_{0}|E|^2$.  

The spontaneous emission driven by $S_{F}$ and $S_{R}$ as a Langevin noise source is performed by Gaussian white noise randomly generated in space and time \cite{299461}, which satisfies the correlation relation,
\begin{equation}
\langle S (t,z) S^* (t',z') \rangle = \frac{2hv}{\varepsilon_0} \Gamma \beta K_p F_p R_{sp} \delta(t - t') \delta(z - z') / v_g
\label{eq:17} 
\end{equation}
\begin{equation}
\langle S (t,z) S (t',z') \rangle = 0
\label{eq:18} 
\end{equation}
where $R_{sp}=N_c/(\tau_{sp} L_c)$ represents the amount of spontaneous emission per unit length of the active region, $\beta$ is the spontaneous emission factor, $\tau_{sp}$ denotes the spontaneous emission lifetime, $K_p$ is the transverse Petermann factor with a fixed value of 1, and $\delta$ is the unit impulse response function. Petermann has proved that spontaneous emission fields coupled to forward and reverse waves have equal amplitudes \cite{1070064}, $S_{F} (t,z)=S_{R} (t,z)=S(t,z)$. The boundary conditions are given by,
\begin{equation}
F(t,0) = r_1 R(t,0), \quad R(t,L_c) = r_2 F(t,L_c)
\label{eq:19} 
\end{equation}
where $r_{1}$ and $r_{2}$ are the wavelength independent mirror reflectivities of PhC cavity’s left and right facets. The boundary of the periodic structure is almost a perfect reflector for Bloch waves, thus reducing loss and increasing the $Q$ factor of the PhC cavity.

\subsection{Carrier rate equation}
A factor $S_{in}$ is added to the pump term to describe the intensity distribution of the pump light along the cavity. Considering Purcell-enhanced spontaneous and stimulated emission, the time-dependent carrier density equation in the active region is described as \cite{Lau:09},
\begin{equation}
\begin{split}
\frac{dN_c(t,z)}{dt} &= \frac{\eta_i L_{in} S_{in}}{hvV} - F_p \beta\frac{ N_c(t,z)}{\tau_{sp}}-(1-\beta)\frac{ N_c(t,z)}{\tau_{sp}}-\frac{N_c(t,z)}{\tau_{nr}} \\
&\quad - v_p 2 g(t,z,\lambda) P(z) F_p N_p (t,z) + D_0 \nabla^2 N_c(t,z)
\end{split}
\label{eq:20}  
\end{equation}
where $\eta_i$ is the internal efficiency, $L_{in}$ denotes the incident pump power, $V$ stands for the volume of active region, $\tau_{nr}$ indicates the non-radiative lifetime, $v_p$ is the phase velocity, and $D_0$ refers to carrier diffusion coefficient \cite{1461503}.

\subsection{Multi-mode rate  equations}
The mode competition process can be described by the multimode rate equations as followed, \cite{1073527,10.1117/12.2035175},
\begin{equation}
\left\{
\begin{aligned}
\frac{1}{v_g}\cdot\frac{dF_m (t,z)}{dt}  + \frac{dF_m (t,z)}{dz} &= \left( \Gamma g(t,z,\lambda) P(z) F_p - \alpha - \frac{\alpha_{MIR}}{S} \right)_m F_m (t,z) + S_{Fm} (t,z) \\
\frac{1}{v_g}  \cdot \frac{dR_m (t,z)}{dt}- \frac{dR_m (t,z)}{dz} &= \left( \Gamma g(t,z,\lambda) P(z) F_p - \alpha - \frac{\alpha_{MIR}}{S}  \right)_m R_m (t,z) + S_{Rm} (t,z) \\
N_{pm} (t,z) &= \frac{|F_m (t,z) + R_m (t,z)|^2 \varepsilon_0}{h v} \\
\frac{dN_c(t,z)}{dt} &= \frac{\eta_i L_{in} S_{in}}{h v V} -F_p \beta \frac{ N_c(t,z)}{\tau_{sp}} - (1-\beta)\frac{ N_c(t,z)}{\tau_{sp}} - \frac{N_c(t,z)}{\tau_{nr}}  \\
&\quad - v_p \sum_m \left( 2 g(t,z,\lambda) P(z) F_p \right)_m N_{pm} (t,z) + D_0 \nabla^2 N_c(t,z)
\end{aligned}
\right.
\label{eq:21}
\end{equation}
where $m$ represents the different Bloch modes. For simplicity, only three Bloch modes are included in the following simulation.

\section{Simulation results and discussion}
In the proposed operation scheme, a cavity $LN$ is formed by removing 20 air holes in a Q1.25 InGaAsP PhC slab with a single InGaAs quantum well in the middle, which emits 1550 nm light in the optical communication C band. Using a cylindrical lens followed by a microscopic lens subsequently arranged on the optical axis, a circular beam of pumping light is transformed into a bar-shaped spot, providing uniform pumping throughout the line-defect cavity. Two parallel beams originally split from the same coherent laser source are directed into the microscopic lens, converged at the focal plane, and superimposed to form an interference fringe. By adjusting the spacing between the parallel beams to modify the incidence angle, the interference fringe pattern can be well aligned with the spatial distribution of the optical mode along the cavity. However, due to the diffraction limit of optical instruments, it is hardly possible to spatially modulate the pumping light with a resolution beyond the lattice constant, making it unrealistic to achieve the full manipulation described by both sinusoid components in Equation \eqref{eq:10}. However, by overlooking the fast-varying component, it is possible to achieve effective single-mode lasing by modulating the pump light only with the slow-varying envelope which corresponds to Bloch modes. This operation requires lower spatial resolution, which is comparable to the cavity length and hence is experimentally viable. 

Based on the above principles, we solve Equation \eqref{eq:21} using a finite differential method by dividing the time and space axes into small segments. The initial condition is set to $N_c={10^{17}} m^{-3}$, $N_{pm}={10^{10}} {m^{-3}}$. The forward and backward fields are initially set to zero and simultaneously begin to propagate in opposite directions from the two facets of the cavity. The forward light field $F_m(t-\Delta t,z)$, which is located at position $z$ at time $t-\Delta t$, propagates to $z+\Delta z$ after $\Delta t$ time and evolves into $F_m(t,z+\Delta z)$. Similarly, the reverse light field $R_m(t-\Delta t,z+\Delta z)$ propagates to position $z$ after $\Delta t$ moments and becomes $R_m(t,z)$. 

The parameters used in the simulation are shown in Table \ref{tab:shape-functions}. The total cavity length is divided into 600 segments in the calculation to provide a spatial resolution much higher than the lasing wavelength. The time and space steps satisfy the stability condition $\Delta t\leq L_c/(600 v_g)$. Following the finite differential method, we simulate the mode characterization of line-defect cavity lasers for two specific cases, under uniform pumping with $S_{in}=1$ and under interference pumping following Equation \eqref{eq:9}. Three lasing modes in the cavity are included in the simulation at the lasing wavelengths of $\lambda_{1}$=1536.7 nm,  $\lambda_{2}$=1539.8 nm and $\lambda_{3}$=1542.8 nm, which correspond to the number of maxima for each Bloch mode $m_{1}=3$, $m_{2}=2$ and $m_{3}=1$, respectively. A continuous wave laser with a wavelength of 976 nm is used as the pump source. The laser incidence angle is set to 9.8 °, 6.5 ° and 3.3 ° to match the three interference pumping conditions and corresponding fringe periods are $L_c/3$, $L_c/2$, and $L_c$.
\begin{table}[htbp]
\caption{Model Simulation Parameters} 
  \label{tab:shape-functions}
  \centering
\begin{tabular}{ccc}
\hline
Parameter & Symbol & Value \\
\hline
Lattice constant                            & $a$            & 430 nm         \\ \hline
Number of defects                           & LN             & 20             \\ \hline
Length of active region                     & $L_c$            & 8.6 $\mu$m         \\ \hline
Width of active region                      & $W$            & 2 $\mu$m           \\ \hline
Thickness of active region                  & $D$            & 5 nm \cite{Wang:22}       \\ \hline
{Slab thickness}                              & ${h_s}$            &{220 nm \cite{Wang:22}} \\ \hline
Volume of active region                       & $V_a$          & 0.086 $\mu$m$^3$   \\ \hline
Optical confinement factor                  & $\Gamma$       & 0.022          \\ \hline
Mode volume                                 & $V_p$          & 3.91 $\mu$m$^3$    \\ \hline
Quality factor                              & $Q$            & 15000          \\ \hline
Internal quantum efficiency                 & $\eta_i$       & 0.2            \\ \hline
Gain compressive factor                     & $\varepsilon$  & $3 \times 10^{-23}$ m$^3$ \cite{299461} \\ \hline
Non-radiative lifetime                      & $\tau_{sp}$    & 2 ns \cite{6471172}      \\ \hline
Spontaneous emission lifetime               & $\tau_{non}$   & 10 ns \cite{6471172}     \\ \hline
Transparent carrier density                 & $N_0$          & $1 \times 10^{23}$ $m^{-3}$ \\ \hline
Internal optical loss                       & $\alpha$       & 800 m$^{-1}$   \\ \hline
Differential gain                           & $g_N$          & $3 \times 10^{-20}$ m$^2$ \cite{299461} \\ \hline
Carrier diffusion coefficient               & $D_0$          & $2 \times 10^{-4}$ m$^2$ s$^{-1}$ \cite{1461503} \\ \hline
Parabolic gain fitting factor               & $G_0$          & $1.2 \times 10^{-9}$ m$^{3/2}$ \\ \hline
Group index                                 & $n_g$          & 20 \cite{PhysRevLett.116.063901}        \\ \hline
Effective refractive index                                 & $n$          & 3.2         \\ \hline
{Effective spontaneous emission factor}                 & $\beta$        & 0.05 \cite{PhysRevLett.116.063901}\\ \hline
{Purcell factor}                              & ${F_p}$          & {$ \sim32$ \cite{Canet-Ferrer:12}}       

 \\ 
\hline
\end{tabular}
\end{table}

\begin{figure}[h]
\centering\includegraphics[width=10.5cm]{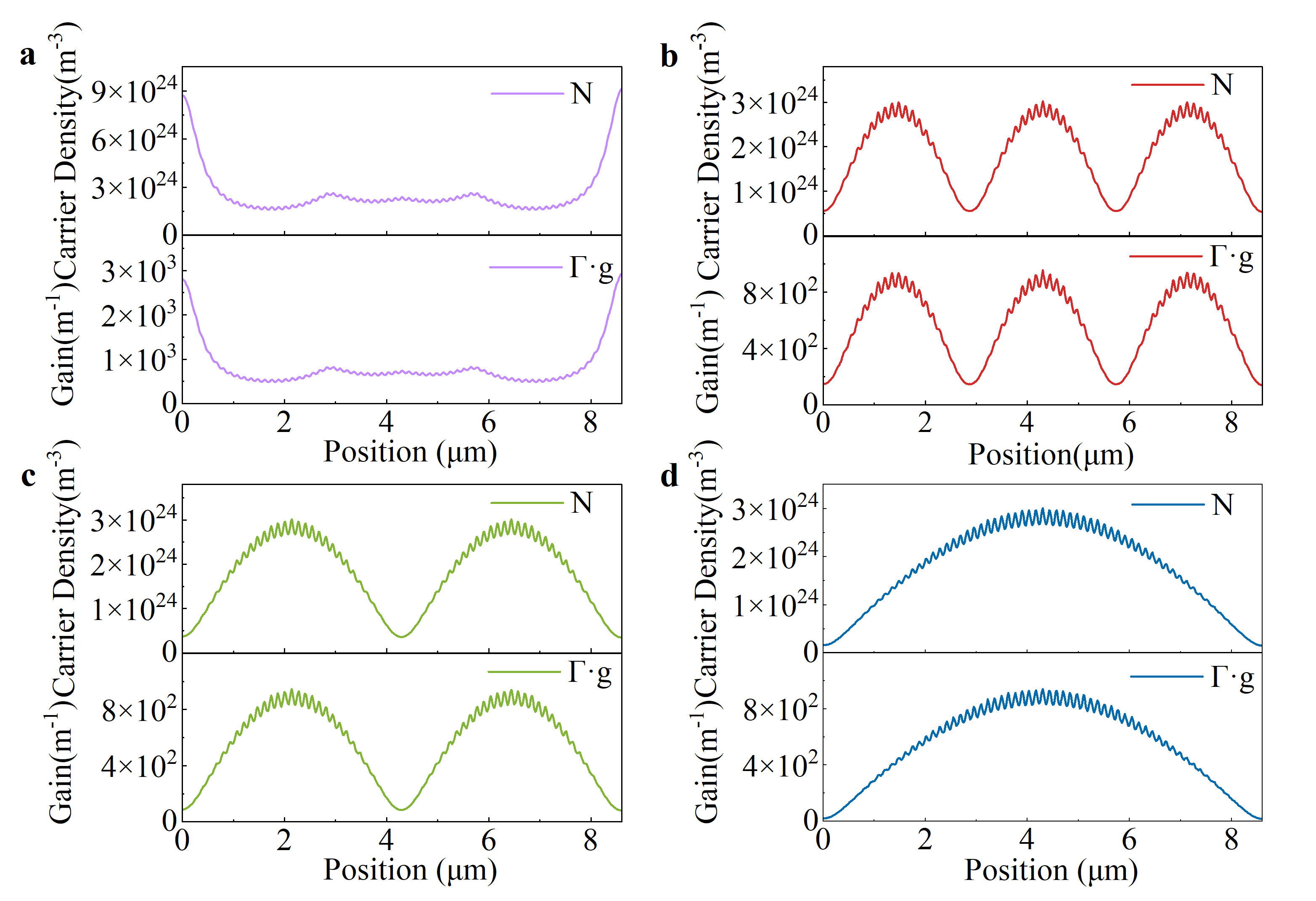}
\caption{The distribution of carrier density and optical gain along the longitudinal direction under the various types of spatial pumping. a. Carrier density and gain distribution under uniform pumping. b. Carrier density and gain distribution under the interference pumping with three interference maxima. c. Carrier density and gain distribution under the interference pumping with two interference maxima. d. Carrier density and gain distribution under the interference pumping with one interference maximum.}
\label{fig:3}
\end{figure}

To study the underlying mechanism for the selection of Bloch modes, the distribution of carrier density and optical gain along the line-defect PhC cavity is calculated. Under uniform pumping, carriers exhibit severe spatial hole burning, as all Bloch modes exhibit nodes at the cavity facets. The field intensity of three Bloch modes on the facet is equal to zero and consumes few carriers, as shown in Fig. \ref{fig:3}(a). When using interference pumping, the pump light is modulated with 3, 2 and 1 maxima along the longitudinal direction of the cavity, which matches well with the mode field of $\lambda_{1}$, $\lambda_{2}$ and $\lambda_{3}$, respectively. The distribution of carriers and optical gain follows the profile of spatial pumping, with 3, 2 and 1 maxima as shown in Fig. \ref{fig:3}(b-d). Spatial pumping leads to the modulation of the imaginary part of the refractive index, forming a gain grating similar to that of gain-coupled DFB lasers \cite{Zhu:15,10.1063/1.1661499}. Apart from the slow-varying envelope of the Bloch modes, very small ripples originating from the fast-varying component with period 2$a$ are observed. The amplitude of those ripples tends to flatten out as a result of carrier diffusion effects. Under spatial injection, the carrier diffusion coefficient has a non-negligible influence on mode selection, which tends to favor our Bloch-mode-selection approach due to the loss of the carrier-density contrast on the fast-varying component.

\begin{figure}[h]
\centering\includegraphics[width=10.5cm]{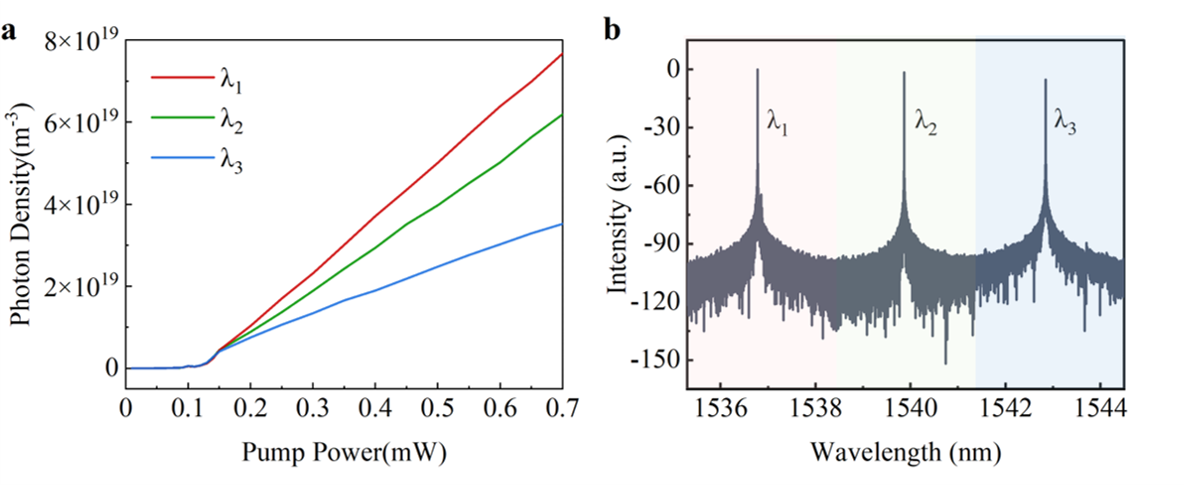}
\caption{$LN$ cavity's lasing performance under uniform optical pumping. a. Light-light curves for three Bloch modes. b. Lasing spectra for the three Bloch modes.}
\label{fig:4}
\end{figure}

\begin{figure}[h]
\centering\includegraphics[width=11cm]{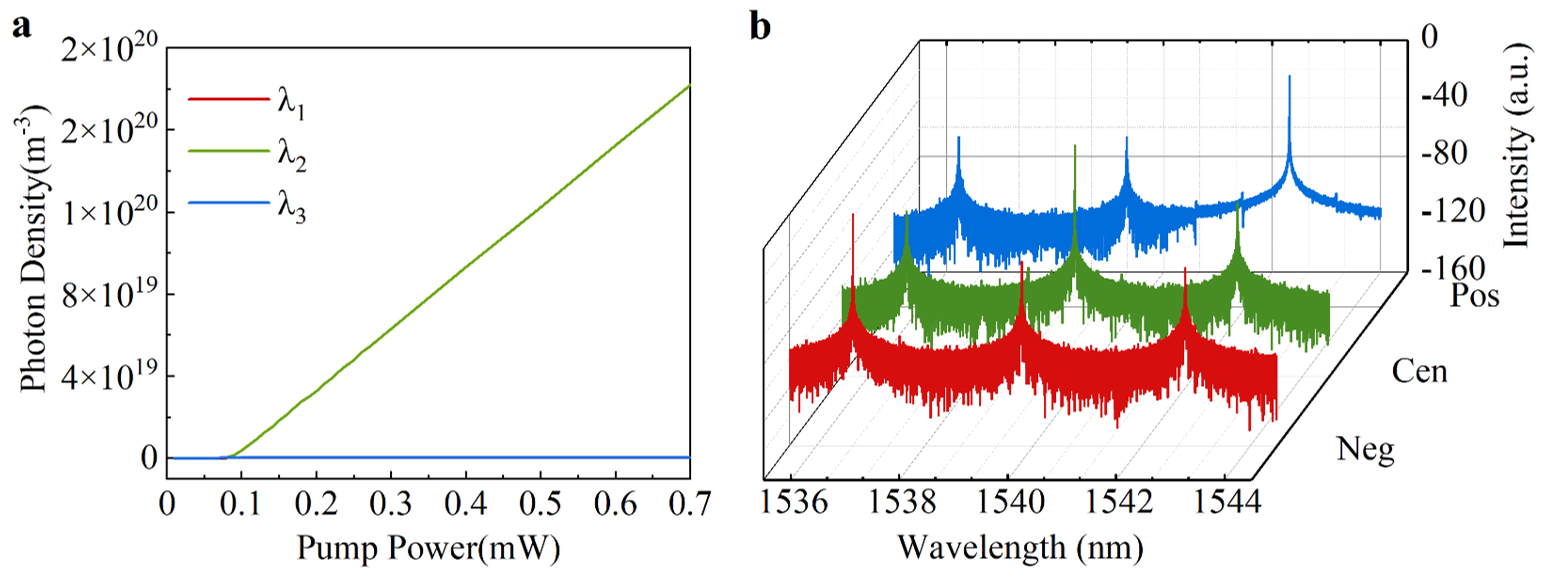}
\caption{Mode selection performance. a. Light-light curves of the three Bloch modes under two-node interference pumping. b. Spectra under  optical interference pumping. The number of envelop maxima for spatial pumping are 1 (POS, blue lines), 2 (Cen, green lines), and 3 (Neg, red lines), respectively.}
\label{fig:5}
\end{figure}

We explore the lasing characteristic by altering the pumping method. Fig. \ref{fig:4}(a) shows the L-L curves for three modes under uniform pumping. The SMSR is 1.5 dB between $\lambda_{1}$ and $\lambda_{2}$, and 3.5 dB between $\lambda_{2}$ and $\lambda_{3}$ at a pumping power of 3 mW, indicating a typical multimode laser as shown in Fig. \ref{fig:4}(b). The threshold power under uniform pumping is around 130 $\mathrm{\mu W}$ for $\lambda_2$. When optical interference injection effectively aligns with the Bloch mode $\lambda_{2}$ field distribution, the threshold power decreases to about 80 $\mu$W as shown in Fig. \ref{fig:5}(a). A distinct shift from multimode to single-mode lasing is visible by comparing Fig. \ref{fig:5}(a) with Fig. \ref{fig:4}(a). Figure \ref{fig:5}(b) presents the output power spectra of three modes at the same pumping power as that under two-maxima interference pumping when Bloch mode $\lambda_{2}$ is selected, with a SMSR of 46.66 dB between $\lambda_{1}$ and $\lambda_{2}$, and 40.27 dB between $\lambda_{2}$ and $\lambda_{3}$. Subsequently, to successfully overlap the field of other Bloch modes, the number of interference nodes is adjusted. For the mode $\lambda_{1}$ (Neg, m=3), as shown in Fig. \ref{fig:5}(b), an effective mode selection is achieved with an SMSR = 33.19 dB to $\lambda_{2}$ and SMSR = 37.76 dB to $\lambda_{3}$. The interference pump with three maxima is tested to select $\lambda_{3}$ (Pos1, m=1). The corresponding SMSR for $\lambda_{1}$ is 42.74 dB and for $\lambda_{2}$ is 43.02 dB. Interference pumping allows for effective mode selection by making the spatial distribution of optical gain overlap the vacuum field of various modes.

\begin{figure}[htbp]
\centering\includegraphics[width=11cm]{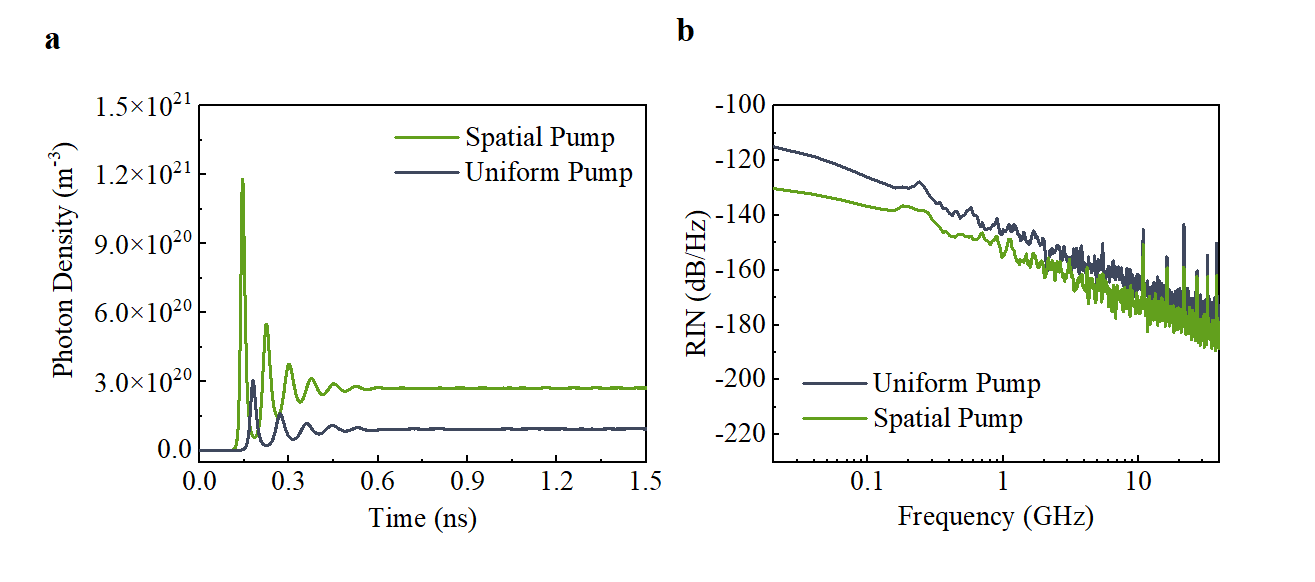}
\caption{Relative intensity noise performance when altering the pumping condition. a. The variation of photon density of the mode $\lambda_2$ over time. b. The relative intensity noise spectra of the central mode.}
\label{fig:6}
\end{figure}

We further study the spectral linewidth and noise characteristics of single-mode lasing via Bloch-mode selection. As the intensity noise results from random fluctuations in photon density over time due to spontaneous emission coupled into the lasing mode \cite{970907}, relative intensity noise (RIN) spectra are defined as the fast Fourier transform of the autocorrelation function that follows Equations \eqref{eq:17} and \eqref{eq:18},

\begin{equation}
\text{RIN} = \frac{1}{P^2} \left\{\left| 
\int_0^T \delta P(t) \, e^{-j\omega t} \, dt 
\right|^2\right\},
\label{eq:22} 
\end{equation}
where $P$ is the output power, $T$ is the total time duration, $dt$ is the time step. Figure \ref{fig:6}(a) shows the time evolution of the photon density where relatively stable emission is obtained after a turn-on delay and relaxation oscillation of the photon density. Figure \ref{fig:6}(b) represents RIN spectra for the mode $\lambda_2$ in uniformly and spatially pumped cavities, which are calculated from the stable part of the time evolution of photon density. Under the same pumping power, the SMSR significantly improves under interference pumping. Photons generated inside the cavity are concentrated in the selected mode, with a density of approximately $2.7\times10^{20} m^{-3}$. In contrast, uniform pumping leads to severe mode competition, with only around $9.4\times10^{19} m^{-3}$ for the photon density in mode $\lambda_2$. The increased proportion of coherent photons leads to the suppression of intensity fluctuation. The overall RIN is significantly reduced under interference pumping, as shown in Fig. \ref{fig:6}(b), because a smaller portion of injection under interference pumping is converted into spontaneous emission coupled to the lasing mode than that under uniform pumping.

\begin{figure}[h]
\centering\includegraphics[width=11cm]{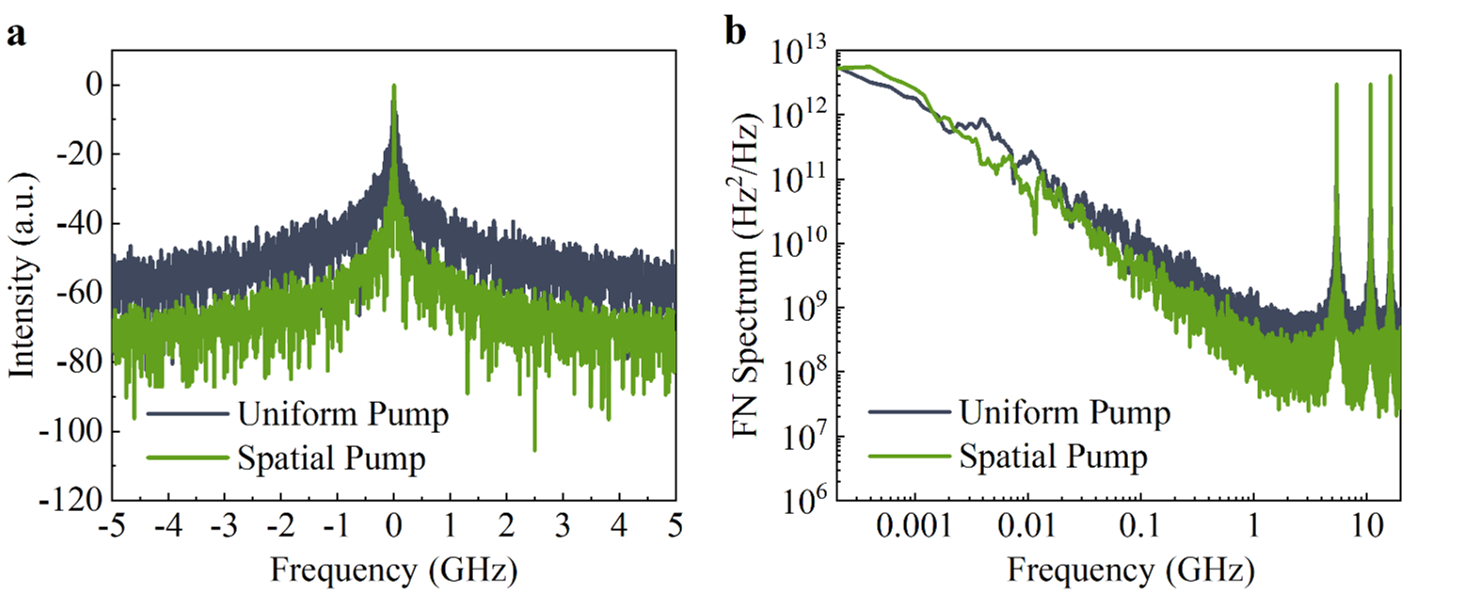}
\caption{Frequency noise performance when altering the pumping condition. a. The lasing spectra under uniform pumping (dark blue line) and interference pumping (green line), indicating a spectral linewidth narrowing via the Bloch mode selection. b. Phase noise curve of the center mode under different pumping conditions.}
\label{fig:7}
\end{figure}

To investigate the effect of interference pumping on phase noise properties, Figure \ref{fig:7}(a) shows the corresponding laser spectra obtained by performing a Fourier transform on the central mode optical field over 0.2 $\mu$s after the relaxation oscillation under both uniform and interference pumping conditions. Figure \ref{fig:7}(b) shows the phase noise spectra defined by the following equation, 
\begin{equation}
FN = \frac{1}{T}\left| 
\int_0^T \frac{1}{2\pi}\frac{d\varphi}{dt} e^{-j\omega t} \, dt 
\right|.
\label{eq:23} 
\end{equation}
The simulation results show that the linewidth of the selected mode under interference pumping is narrower than that under uniform pumping, implying a lower proportion of spontaneous emission coupled into the lasing mode, which improves the noise characteristics. Under uniform pumping, photons coupled into the lasing mode from spontaneous emission occur at each position in the longitudinal direction. The distribution of the number of these photons becomes disordered when compared to the stimulated emission of the lasing mode which originally has the same node distribution as the vacuum field. The spontaneous emission under uniform pumping appears as random radiation sources distributed across all frequencies \cite{PhysRev.93.99,e24121717}. In contrast, when the optical pumping forms interference fringes, the carrier injection tends to favor the mode with similar field distribution. The amount of spontaneous emission coupled into the selected Bloch mode is reduced, which suppresses the noise in the lasing mode. Meanwhile, the threshold of the selected lasing mode is reduced, further narrowing the linewidth \cite{8038042}. 

It is evident that the lasing characteristics under interference pumping are improved with higher SMSR, narrower linewidth and lower noise. Strictly speaking, ignoring the fast-varying component of lasing mode, the effectiveness of mode manipulation might be downgraded. However, because of the carrier diffusion at small distance, the modulation of optical gain through the slow-varying component can be still effective and maintains the overall performance. The ultimate effectiveness of this Bloch-mode-selection approach depends on the overlap between the gain distribution after carrier diffusion and the vacuum field distribution inside nanophotonic structures.

\section{Conclusion}
In conclusion, this work proposes a spatial pumping scheme to realize single-mode lasing in PhC lasers via Bloch mode selection. When the spatial profile of optical interference injection overlaps with the slow-varying envelope of a Bloch mode, single mode lasing with a high SMSR of 30 dB is numerically observed with narrowed lasing linewith and improved noise characteristics. While introducing the quantum-optics understanding of vacuum field into a semiclassical model based on traveling wave rate equations, the approach of Bloch mode selection demonstrates itself as a novel and viable tool for the effective mode manipulation of semiconductor lasers using periodical structures, which can generate far-reaching impacts in the field of nanophotonics using PhCs and topological structures.

\begin{backmatter}
\bmsection{Funding}\\
This work is supported by the National Key Research and Development Program of China (2021YEB2800500), the National Natural Science Foundation of China (61574138, 61974131), and the Zhejiang Province Leading Geese Plan (2024C01105).


\bmsection{Disclosures}\\
The authors declare no conflicts of interest.
\end{backmatter}

\bibliography{sample}

\begin{thebibliography}{10}
\newcommand{\enquote}[1]{``#1''}

\bibitem{Shi:22}
Y.~Shi, Y.~Zhang, Y.~Wan, \emph{et~al.}, \enquote{Silicon photonics for high-capacity data communications,} {\protect\JournalTitle{Photon. Res.}} \textbf{10}, A106--A134 (2022).

\bibitem{Wang2020b}
W.~Jianwei, S.~Fabio, L.~Anthony, \emph{et~al.}, \enquote{Integrated photonic quantum technologies,} {\protect\JournalTitle{Nature Photonics}} \textbf{14}, 273--284 (2020).

\bibitem{Takeda:21}
K.~Takeda, T.~Tsurugaya, T.~Fujii, \emph{et~al.}, \enquote{Optical links on silicon photonic chips using ultralow-power consumption photonic-crystal lasers,} {\protect\JournalTitle{Opt. Express}} \textbf{29}, 26082--26092 (2021).

\bibitem{10.1117/1.AP.1.1.014002}
C.~Z. Ning, \enquote{{Semiconductor nanolasers and the size-energy-efficiency challenge: a review},} {\protect\JournalTitle{Advanced Photonics}} \textbf{1}, 014002 (2019).

\bibitem{https://doi.org/10.1002/lpor.202400218}
X.~T. Cheng, L.~F. Wang, Y.~Z. Li, \emph{et~al.}, \enquote{Topologically protected single edge mode lasing in photonic crystal su–schrieffer–heeger lattice with directional loss control,} {\protect\JournalTitle{Laser \& Photonics Reviews}} \textbf{18}, 2400218 (2024).

\bibitem{10.1063/1.2716972}
H.~Hofmann, H.~Scherer, S.~Deubert, \emph{et~al.}, \enquote{Spectral and spatial single mode emission from a photonic crystal distributed feedback laser,} {\protect\JournalTitle{Applied Physics Letters}} \textbf{90}, 121135 (2007).

\bibitem{10.1063/1.1427158}
T.~D. Happ, A.~Markard, M.~Kamp, \emph{et~al.}, \enquote{Single-mode operation of coupled-cavity lasers based on two-dimensional photonic crystals,} {\protect\JournalTitle{Applied Physics Letters}} \textbf{79}, 4091--4093 (2001).

\bibitem{Wang:22}
L.~F. Wang, X.~T. Cheng, X.~D. Zhang, \emph{et~al.}, \enquote{Mode selection in {InGaAs/InGaAsP} quantum well photonic crystal lasers based on coupled double-heterostructure cavities,} {\protect\JournalTitle{Opt. Express}} \textbf{30}, 10229--10238 (2022).

\bibitem{Burgwal2023}
R.~Burgwal and E.~Verhagen, \enquote{Enhanced nonlinear optomechanics in a coupled-mode photonic crystal device,} {\protect\JournalTitle{Nature Communications}} \textbf{14}, 1526 (2023).

\bibitem{PhysRevA.91.063807}
R.~Johne, R.~Schutjens, S.~Fattah~poor, \emph{et~al.}, \enquote{Control of the electromagnetic environment of a quantum emitter by shaping the vacuum field in a coupled-cavity system,} {\protect\JournalTitle{Phys. Rev. A}} \textbf{91}, 063807 (2015).

\bibitem{Wang:21}
L.~F. Wang, X.~T. Cheng, X.~D. Zhang, \emph{et~al.}, \enquote{Mode selection in {L}40 photonic crystal cavities via spatially distributed pumping,} in \emph{Asia Communications and Photonics Conference,}  (Optica Publishing Group, 2021), pp. T4A--198.

\bibitem{Kim2020}
H.~R. Kim, M.~S. Hwang, D.~Smirnova, \emph{et~al.}, \enquote{Multipolar lasing modes from topological corner states,} {\protect\JournalTitle{Nature Communications}} \textbf{11}, 5758 (2020).

\bibitem{Wang:20}
L.~F. Wang, Y.~R. Wang, H.~Francis, \emph{et~al.}, \enquote{Theoretical modelling of single-mode lasing in microcavity lasers via optical interference injection,} {\protect\JournalTitle{Opt. Express}} \textbf{28}, 16486--16496 (2020).

\bibitem{Gu2017}
F.~Gu, F.~Xie, X.~Lin, \emph{et~al.}, \enquote{Single whispering-gallery mode lasing in polymer bottle microresonators via spatial pump engineering,} {\protect\JournalTitle{Light: Science \& Applications}} \textbf{6}, e17061 (2017).

\bibitem{10.1063/1.5012112}
B.~Rigal, D.~Drahi, C.~Jarlov, \emph{et~al.}, \enquote{Probing disorder and mode localization in photonic crystal cavities using site-controlled quantum dots,} {\protect\JournalTitle{Journal of Applied Physics}} \textbf{123}, 043109 (2018).

\bibitem{Cheng2024QuantumVacuum}
X.~Cheng, L.~Wang, J.~Yu, \emph{et~al.}, \enquote{Manipulation of quantum vacuum field for microcavity photonics (invited),} {\protect\JournalTitle{ACTA PHOTONICA SINICA}} \textbf{53}, 0553104 (2024).

\bibitem{Okano_2010}
M.~Okano, T.~Yamada, J.~Sugisaka, \emph{et~al.}, \enquote{Analysis of two-dimensional photonic crystal l-type cavities with low-refractive-index material cladding,} {\protect\JournalTitle{Journal of Optics}} \textbf{12}, 075101 (2010).

\bibitem{Vasco2018}
J.~P. Vasco and S.~Hughes, \enquote{Anderson localization in disordered ln photonic crystal slab cavities,} {\protect\JournalTitle{ACS Photonics}} \textbf{5}, 1262--1272 (2018).

\bibitem{saleh2019fundamentals}
B.~E. Saleh and M.~C. Teich, \emph{Fundamentals of photonics} (john Wiley \& sons, 2019).

\bibitem{lalanne2008photon}
P.~Lalanne, C.~Sauvan, and J.~P. Hugonin, \enquote{Photon confinement in photonic crystal nanocavities,} {\protect\JournalTitle{Laser \& Photonics Reviews}} \textbf{2}, 514--526 (2008).

\bibitem{nano11113030}
M.~Saldutti, M.~Xiong, E.~Dimopoulos, \emph{et~al.}, \enquote{Modal properties of photonic crystal cavities and applications to lasers,} {\protect\JournalTitle{Nanomaterials}} \textbf{11}, 3030 (2021).

\bibitem{1291710}
C.~Y. Jin, Y.~Z. Huang, L.~J. Yu, \emph{et~al.}, \enquote{Detailed model and investigation of gain saturation and carrier spatial hole burning for a semiconductor optical amplifier with gain clamping by a vertical laser field,} {\protect\JournalTitle{IEEE Journal of Quantum Electronics}} \textbf{40}, 513--518 (2004).

\bibitem{Hantschmann:18}
C.~Hantschmann, P.~P. Vasil'ev, S.~Chen, \emph{et~al.}, \enquote{Gain switching of monolithic 1.3 $\mu$m inas/gaas quantum dot lasers on silicon,} {\protect\JournalTitle{J. Lightwave Technol.}} \textbf{36}, 3837--3842 (2018).

\bibitem{1291704}
W.~Li, W.-P. Huang, and X.~Li, \enquote{Digital filter approach for simulation of a complex integrated laser diode based on the traveling-wave model,} {\protect\JournalTitle{IEEE Journal of Quantum Electronics}} \textbf{40}, 473--480 (2004).

\bibitem{PhysRevLett.116.063901}
W.~Xue, Y.~Yu, L.~Ottaviano, \emph{et~al.}, \enquote{Threshold characteristics of slow-light photonic crystal lasers,} {\protect\JournalTitle{Phys. Rev. Lett.}} \textbf{116}, 063901 (2016).

\bibitem{10.1143/ptp/5.4.570}
E.~Fermi, \enquote{High energy nuclear events,} {\protect\JournalTitle{Progress of Theoretical Physics}} \textbf{5}, 570--583 (1950).

\bibitem{Jin2014}
C.-Y. Jin, R.~Johne, M.~Y. Swinkels, \emph{et~al.}, \enquote{Ultrafast non-local control of spontaneous emission,} {\protect\JournalTitle{Nature Nanotechnology}} \textbf{9}, 886--890 (2014).

\bibitem{10.1063/1.3697702}
N.~Gregersen, T.~Suhr, M.~Lorke, \emph{et~al.}, \enquote{Quantum-dot nano-cavity lasers with purcell-enhanced stimulated emission,} {\protect\JournalTitle{Applied Physics Letters}} \textbf{100}, 131107 (2012).

\bibitem{Zhou2020}
T.~Zhou, M.~Tang, G.~Xiang, \emph{et~al.}, \enquote{Continuous-wave quantum dot photonic crystal lasers grown on on-axis si (001),} {\protect\JournalTitle{Nature Communications}} \textbf{11}, 977 (2020).

\bibitem{Purcell1995}
E.~M. Purcell, \enquote{Spontaneous emission probabilities at radio frequencies,} in \emph{Confined electrons and photons: new physics and applications,}  (Springer, 1995), pp. 839--839.

\bibitem{1071656}
T.-P. Lee, C.~Burrus, J.~Copeland, \emph{et~al.}, \enquote{Short-cavity ingaasp injection lasers: Dependence of mode spectra and single-longitudinal-mode power on cavity length,} {\protect\JournalTitle{IEEE Journal of Quantum Electronics}} \textbf{18}, 1101--1113 (1982).

\bibitem{299461}
L.~Zhang, S.~Yu, M.~Nowell, \emph{et~al.}, \enquote{Dynamic analysis of radiation and side-mode suppression in a second-order dfb laser using time-domain large-signal traveling wave model,} {\protect\JournalTitle{IEEE Journal of Quantum Electronics}} \textbf{30}, 1389--1395 (1994).

\bibitem{1070064}
K.~Petermann, \enquote{Calculated spontaneous emission factor for double-heterostructure injection lasers with gain-induced waveguiding,} {\protect\JournalTitle{IEEE Journal of Quantum Electronics}} \textbf{15}, 566--570 (1979).

\bibitem{Lau:09}
E.~K. Lau, A.~Lakhani, R.~S. Tucker, and M.~C. Wu, \enquote{Enhanced modulation bandwidth of nanocavity light emitting devices,} {\protect\JournalTitle{Opt. Express}} \textbf{17}, 7790--7799 (2009).

\bibitem{1461503}
K.~Nozaki and T.~Baba, \enquote{Carrier and photon analyses of photonic microlasers by two-dimensional rate equations,} {\protect\JournalTitle{IEEE Journal on Selected Areas in Communications}} \textbf{23}, 1411--1417 (2005).

\bibitem{1073527}
R.~Olshansky, P.~Hill, V.~Lanzisera, \emph{et~al.}, \enquote{Frequency response of 1.3µm {InGaAsP} high speed semiconductor lasers,} {\protect\JournalTitle{IEEE Journal of Quantum Electronics}} \textbf{23}, 1410--1418 (1987).

\bibitem{10.1117/12.2035175}
M.~Yousefi and D.~Lenstra, \enquote{Rate-equation description of multi-mode semiconductor lasers,} in \emph{Physics and Simulation of Optoelectronic Devices XXII,}  vol. 8980 (SPIE, 2014), pp. 57--75.

\bibitem{6471172}
S.~Matsuo, T.~Sato, K.~Takeda, \emph{et~al.}, \enquote{Ultralow operating energy electrically driven photonic crystal lasers,} {\protect\JournalTitle{IEEE Journal of Selected Topics in Quantum Electronics}} \textbf{19}, 4900311--4900311 (2013).

\bibitem{Canet-Ferrer:12}
J.~Canet-Ferrer, L.~J. Mart\'{i}nez, I.~Prieto, \emph{et~al.}, \enquote{Purcell effect in photonic crystal microcavities embedding inas/inp quantum wires,} {\protect\JournalTitle{Opt. Express}} \textbf{20}, 7901--7914 (2012).

\bibitem{Zhu:15}
H.~Zhu, Y.~Xia, and J.-J. He, \enquote{Pattern dependence in high-speed q-modulated distributed feedback laser,} {\protect\JournalTitle{Opt. Express}} \textbf{23}, 11887--11897 (2015).

\bibitem{10.1063/1.1661499}
H.~Kogelnik and C.~V. Shank, \enquote{Coupled‐wave theory of distributed feedback lasers,} {\protect\JournalTitle{Journal of Applied Physics}} \textbf{43}, 2327--2335 (1972).

\bibitem{970907}
M.~Ahmed, M.~Yamada, and M.~Saito, \enquote{Numerical modeling of intensity and phase noise in semiconductor lasers,} {\protect\JournalTitle{IEEE Journal of Quantum Electronics}} \textbf{37}, 1600--1610 (2001).

\bibitem{PhysRev.93.99}
R.~H. Dicke, \enquote{Coherence in spontaneous radiation processes,} {\protect\JournalTitle{Phys. Rev.}} \textbf{93}, 99--110 (1954).

\bibitem{e24121717}
A.~M. Cetto and L.~de~la Pe{\~n}a, \enquote{The electromagnetic vacuum field as an essential hidden ingredient of the quantum-mechanical ontology,} {\protect\JournalTitle{Entropy}} \textbf{24}, 1717 (2022).

\bibitem{8038042}
H.~Z. Weng, Y.~Z. Huang, X.~W. Ma, \emph{et~al.}, \enquote{Spectral linewidth analysis for square microlasers,} {\protect\JournalTitle{IEEE Photonics Technology Letters}} \textbf{29}, 1931--1934 (2017).

\end{thebibliography}





\end{document}